# Return level estimations for extreme rainfall over the Iberian Peninsula: comparing methodologies.

F. Javier Acero [1,2*], Sylvie Parey [3], J. Agustín García[1,2] and Didier Dacunha-Castelle[4]

Dpto. Física, Universidad de Extremadura, Avda. de Elvas s/n, 06006, Badajoz; fjacero@unex.es

[2] Instituto Universitario de Investigación del Agua, Cambio Climático y Sostenibilidad (IACYS), Universidad de Extremadura, E-06006 Badajoz, Spain Badajoz. agustin@unex.es

[3] EDF/R&D, 6 quai Watier, 78401 Chatou Cedex, France; sylvie.parey@edf.fr

[4] Laboratoire de Mathématiques, Université Paris 11, Orsay, France; dacunhacastelledidier@gmail.com

**\*** Correspondence: fjacero@unex.es.

**Abstract:** Different ways to estimate future return levels for extreme rainfall are described and applied to the Iberian Peninsula (IP), based on Extreme Value Theory (EVT). This study is made for an ensemble of high quality rainfall time series observed in the Iberian Peninsula over the period 1961-2010. Both, peaks-over-threshold (POT) approach and block maxima with the Generalized Extreme Value (GEV) distribution will be used and their results compared when linear trends are assumed in the parameters: threshold and scale parameter for POT and location and scale parameter for GEV. Both all-days and rainy-days-only data sets were considered, because rainfall over the IP is a special variable in that a large number of the values are 0. Another methodology is then tested, for rainy days only, considering the role of how the mean, variance, and number of rainy days evolve. The 20-year return levels (RLs) expected in 2020 were estimated using these methodologies for three seasons: fall, spring and winter. GEV is less reliable than POT because fixed blocks lead to the selection of non-extreme values. Future RLs obtained with POT are higher than those estimated with GEV, mainly for some gauges showing significant positive trend for the number of rainy days. Fall becomes the season with heaviest rainfall, rather than winter nowadays, for some regions.

### 1. Introduction

The precipitation regime in the Iberian Peninsula (IP) is highly variable due to its complex topography. leaving no wide areas without coverage given that the spatial variability of IP rainfall is such that certain places receive more than 3000 mm/yr, while others, for example, in the southeast, receive on average less than 200 mm/yr, the lowest values in Europe. The atmospheric circulation patterns over the IP change as a function of the season [1]. Although precipitation is generated by different physical processes at different times of year [2], the intensification of rainfall during the rainy seasons is due to frontal systems coming from the Atlantic Ocean [3, 4], which cause persistent rainfall. For the eastern IP, rainfall is produced by easterly flows and leads to heavy convective precipitation over the Mediterranean area, especially when there is colder air at high levels. Furthermore, over broad regions of the IP, a few rainy days concentrate much of the annual precipitation [5]. This variability leads to an interesting study of extreme rainfall over the IP.

Previous rainfall extreme studies covering particular regions in Spain, Spain or the whole Iberian Peninsula (IP) have considered a variety of indices, but all of them taken from the set of Expert Team on Climate Change Detection and Indices (ETCCDI) indices recommended for use by the World Meteorological Organization [6]. Among others, [7] for Andalusia, the most noteworthy results were found in winter, with a predominance of decreasing trends for intensity indices in western and central (CW) Andalusia, whereas positive trends appeared in the south-eastern (SE) zone, and a study for the whole Iberian Peninsula [8] analyzing annual and seasonal trends of extreme precipitation indices performed using observations for Portugal and Spain and the ERA-driven simulation.

There are some studies of extreme rainfall for the Iberian Peninsula using the statistical extreme value theory (EVT), some of them using the block maxima (BM) approach [9] or using the peaks-over-threshold (POT) technique [10, 11] to study trends in extreme rainfall. But these methodologies have advantages and disadvantages and it is important to compare the results obtained from them. Recently, a new method for the calculation of non-stationary return levels (RLs) for extreme rainfall for a southwestern region of the IP has been developed[12]. The aim in this work is not to propose a methodology for the RLs estimation but to compare two ways of dealing with possible trends in rainfall to estimate near future extremes. This objective is tackled for a set of complete daily rainfall time series from 76 gauges for the period 1961-2010 for the whole Iberian Peninsula.

The organization of the paper is as follows: the data used are described in Section 2, the methodology is lightly described in Section 3 and the main results are presented and discussed in Section 4. Then, the main conclusions are drawn in Section 5.

## 2. Data

The study area was the Iberian Peninsula. Most of the time series were taken from an extensive database of daily rainfall time series provided by the Spanish National Meteorology Agency (AEMET). The quality requirements were to choose time series with no missing data and to cover the orographic diversity of the Iberian Peninsula.

The final choice was a set of 76 daily rainfall time series corresponding to gauges as regularly spaced as possible over the IP. Their locations are shown in Fig. 1. The study period was 1961 to 2010. Most of the time series were selected from the daily rainfall time series of the Spanish National Meteorology Service (AEMET) according to the quality requirements described above. Four of the series were provided and quality controlled by the European Climate Assessment and Dataset service (ECA, available online at http://eca.knmi.nl) and one series was from Portugal (Gafanha da Nazaré ) provided by the National System of Water Resources Information (SNIRH, available online at http://snirh.apambiente.pt/, managed by the Portuguese Institute for Water) database.

Data homogeneity was evaluated using the R-based program RHTestV3, developed at the Climate Research Branch of the Meteorological Service of Canada, and available from the ETCCDMI Web site (http://etccdi.pacificclimate.org/). This program is capable of identifying multiple step changes at documented or undocumented change-points. It is based on a two-phase regression model with a linear trend for the entire base series [13, 14]. This analysis, together with the metadata of the stations, showed that none of the 76 time series had change-points significant at 5%, with all of them being homogeneous in the cited period of study.

For this study of precipitation extremes over the IP, in view of the highly seasonal nature of rainfall, each season was studied separately. The working definition of the seasons was that common in climatological studies: winter was taken to be December, January, and February; spring March, April, and May; and autumn September, October, and November. Summer was not considered because of the low number of rainy events in that season in most parts of the IP. For the BM approach, each season was considered to select its maximum value. For the POT approach, the threshold used for the definition of extreme rainfall, for each season separately, was the 98th percentile of all the daily rainfall values for the whole period, and for the rainy days only, the 95th percentile of the rainfall values.

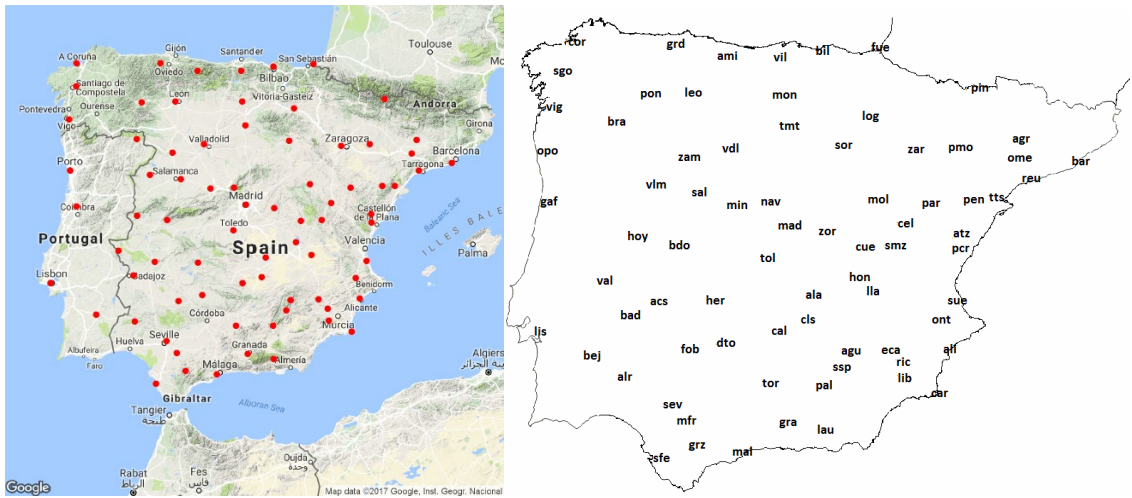

Figure 1. Spatial distribution of the gauges used.

## 3. Methods

The 20-year RLs for the precipitation in the Iberian Peninsula were studied using two different approaches from the EVT: the block maxima (BM) and the peaks-over-threshold (POT) approach, considering both all days (rainy and non-rainy) and rainy days only. The BM approach is frequently applied to the study of extreme events of meteorological variables in the framework of EVT, focused on the Generalized Extreme Value (GEV) distribution which models the block maxima of the variables. GEV theory is a kind of "law of large numbers." It states that for n large enough, the maximum of n independent identically distributed variables with probability distribution function of type F tends to follow a GEV distribution when n tends to infinity.

In a similar way, the POT approach is based on the asymptotical convergence of the exceedances of a high threshold $u$ to a general Pareto distribution (GPD) when $u$ tends to infinity. This holds true again for independent and identically distributed variables.

Then how to make an optimal choice for the threshold $u$ is a difficult problem. A value of $u$ that is too high leads to few exceedances and consequently high variance estimators. On the contrary, a value of $u$ that is too low is likely to violate the asymptotic basis of the model, leading to biases [15].

Independence and identical distribution is an assumption of the theory, but it is generally not verified for meteorological variables. Weak dependence can be dealt with: annual maxima can generally be considered as independent, and the values above the threshold can be made independent, for example by applying a run declustering procedure [16]. Identical distribution is not true because of seasonality, interannual variability and possible trends. To handle seasonality, the estimations are made for each season rather than for the whole year. Interannual variability and trends are more difficult to tackle. The estimation of Return Levels in a non stationary context has lead to an increasing number of papers in the recent years. While some authors consider that taking non stationarity into account implies too large uncertainties to be justified [17, 18], others suggest methodologies to do so together with new definitions for a return level in such a context [19-24].

The aim here is not to derive a Return Level for some design purpose, but to compare two ways of dealing with possible trends in rainfall for the estimation of near future extremes. Thus the focus is not on the definition of a non stationary Return Level, but on the type of trends to consider and the consequences of the choice for both BM and POT approaches.

Thus two different approaches were used to compute future RLs:

1)      The first consists in identifying trends in the parameters of the asymptotic distributions of extremes: location and scale parameter of the GEV for BM, threshold and scale parameter of the GPD for POT. The trend identification is based on likelihood ratio tests with a 5% significance level between models with linear or constant parameters (Coles 2001) and the confidence intervals are computed by bootstrapping in order to take the uncertainty in the trends into account (see the Appendix in [12] for details).

2)      A residual process widely explained in [12] is used in order to calculate the 20-year RLs in 2020. The idea is to use trends in the main characteristics of the whole distribution rather than trends in the extreme values only. Then, if the non-stationarity of extremes is from a statistical framework explained by that of the mean and the variance, the daily mean and standard deviation in 2020 are estimated by linear extrapolation of the linear trends estimated from observations to compute the RLs in 2020.

## 4. Discussion

This section shows the main results of the 20-year RLs (Z20) estimations. Firstly, for present time estimated using both all days (section 4.1) and rainy days only (section 4.2), and secondly, for near future time, i.e, the 20-year RLs in 2020 (section 4.3).

### 4.1. Present 20-year RLs estimated from all days.

Considering all days, rainy and non-rainy, the 20-year RLs for both approaches were estimated for present time from the daily rainfall time series. From the RLs estimates, and analyzing if the Z20-POT is inside the CI of Z20-GEV, figure 2 shows the results for the three seasons considered. For autumn and spring, all the 20-year RLs estimated using POT approach lies inside the CI of the 20-year RLs using GEV. In winter, only one gauge (vlm) shows not overlapping CIs being the RL estimated with POT (55.09[47.96;62.22]) higher than the corresponding with BM (50.16[45.51;54.81]). The difference between both values is low but as the CI with BM is narrower than with POT, the RL estimated with POT does not fall inside GEV CI.

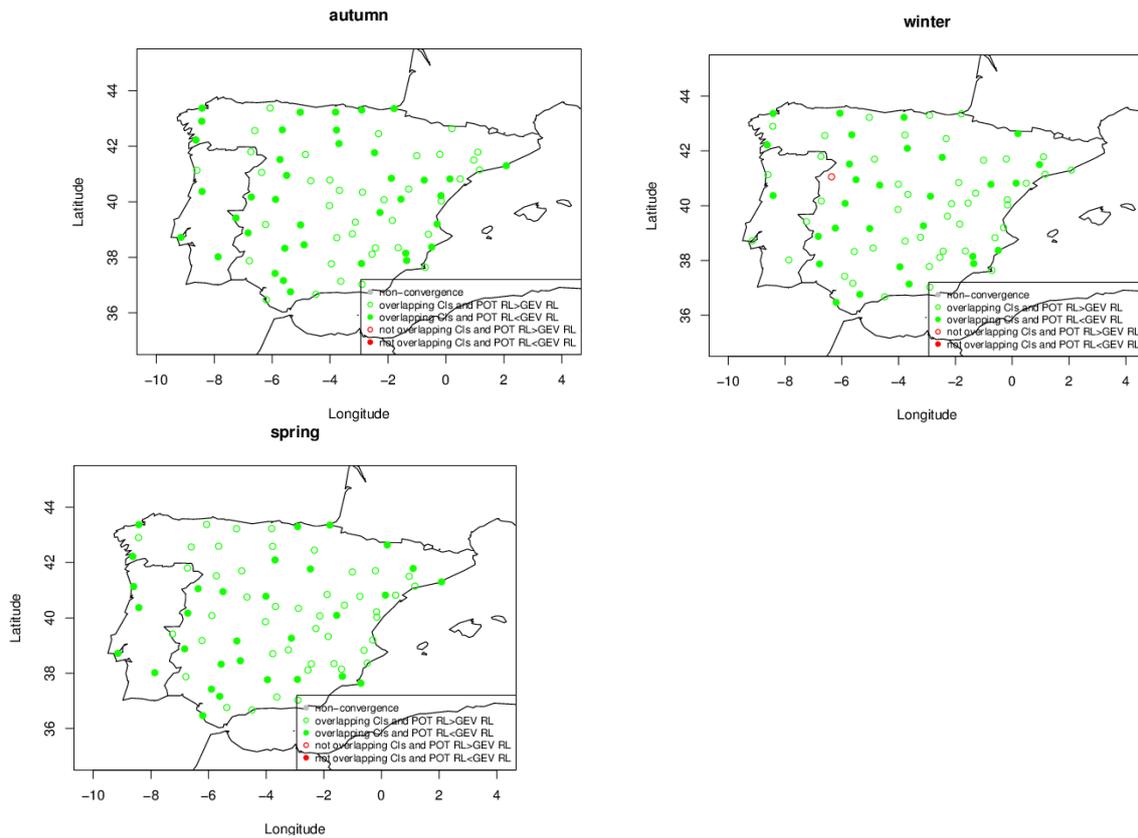

Figure 2. Spatial distribution of 20-year RLs estimated from all days obtained for POT that lie or do not lie inside the CI of the 20-year RLs obtained for BM.

In order to analyze this exception, Figure 3 shows the GEV-POT comparison for this gauge (vlm) in winter. Looking at the top figure, the frequency of threshold exceedances decreases in the last decade and the last values are lower than the rest. But going into detail about the red points for GEV, as one maximum is chosen for each season, high and low values are mixed, some of them being lower than the POT threshold. Also, the shape parameter is different $\xi=0$ for POT and $\xi=-0.2845$ for GEV. In this last case, the distribution is very bounded, which is surprising for precipitation. As sigma measures the variability, the scale parameter is higher in GEV because the variability of the maximum is higher than in POT when choosing so low values as maximum (one season with a rain amount lower than 10 mm is not a really maximum). The estimations are made as if the time series were stationary, which is probably not the case and may further explain the differences.

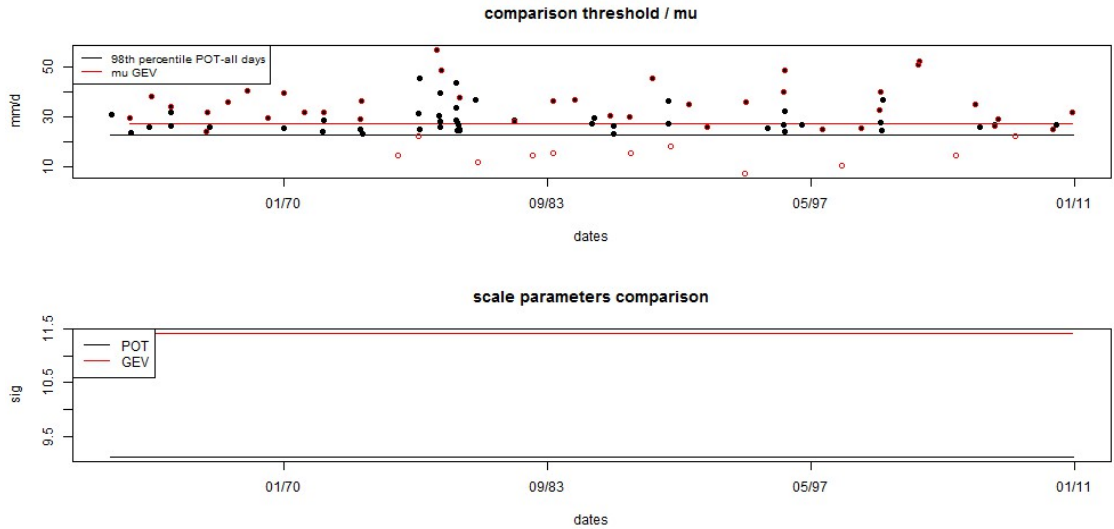

Figure 3.- Comparison GEV-POT for the gauge "vlm" in winter. Top panel shows the comparison between the threshold from POT and the location parameter from BM. Symbols show the independent exceedances for POT (black) and the seasonal maxima from BM (red). Bottom panel shows the scale parameter comparison for both methodologies.

### 4.2. Present 20-year RLs estimated from rainy days.

Considering rainy days only, and studying when Z20-POT is inside the CI of Z20-GEV, figure 4 shows the results for the three seasons considered. For autumn (pcr) and winter (cal), all the gauges show Z20-POT inside the Z20-GEV CI except one in each season; and for spring, there are three (cal, gra, ssp). In all cases, Z20-POT is higher than Z20-GEV. The CI obtained using GEV is narrower than the corresponding with POT.

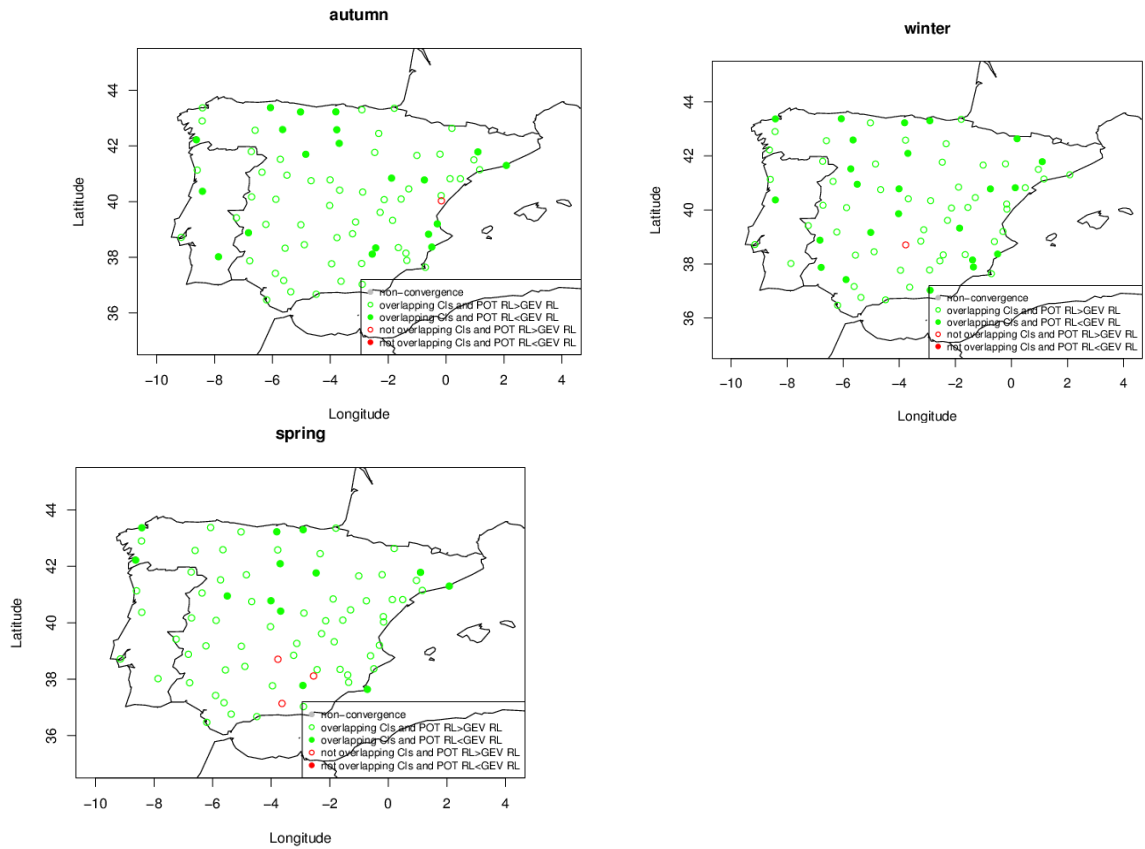

Figure 4.- As Figure 2, but showing the 20-year RLs estimated for rainy days only.

Those gauges showing not overlapping CIs and different values for the RL show the same behavior. Figure 5 shows the GEV-POT comparison for the gauge "pcr" in autumn. There are lots of maximum in the second half of the study period for BM approach with low amount of rainfall leading to lower values of the return levels.

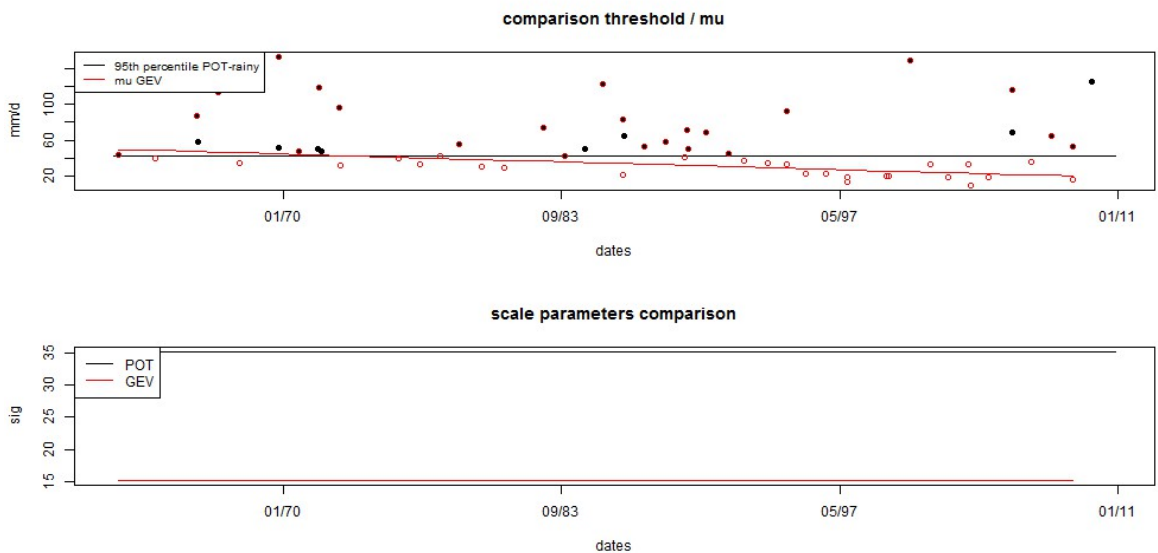

Figure 5.- As Figure 3 but showing GEV-POT comparison for gauge "pcr" in autumn.

Generally, as was expected, both approaches lead to similar results using rainy days only to that in the previous section when using all days, rainy and non-rainy. There were only a few gauges showing not overlapping CIs (1 for autumn and winter and 3 for winter) being the RL obtained from POT approach higher than the corresponding from BM. It is important to remark that because of the amount of zeros, the asymptotic in the case of BM is weakened in rainfall time series, i. e. the maximum is not the maximum of 90 values (which is already quite far from infinity), but of much less.

### 4.3. Future 20-year RLs

The main results of calculating the 20-year RLs in 2020 are presented in this section. Firstly, it is remarkable that, in a stationary context, the value of the shape parameter is zero for most of the stations for the three seasons considered according to the likelihood ratio test at a 95% confidence level.

4.3.1. Using trends in the parameters.

The study of the future 20-year RLs obtained through the extrapolation of the trends in the parameters (scale and location for GEV, and scale and threshold for POT), and looking if the RLs obtained with POT are inside the CI of RLs obtained with GEV, the spatial distribution of the gauges that verify or don't verify this assumption is shown in Figure 6. As can be seen, there are several gauges with not overlapping CIs: 10 for autumn, 11 for winter, and 5 for spring. Looking into deep for these gauges, Tables 2-4 show the degree of each parameter for GEV and POT with 0 meaning no trend and 1 linear trend. Most of these gauges show discrepancies between the degree obtained for the parameters using both methods, when there is a linear trend in scale or location parameters using GEV, there is no trend in scale parameter using POT and vice versa. This leads to different values of the 20-year RL in 2020 using the extrapolation of the parameters.

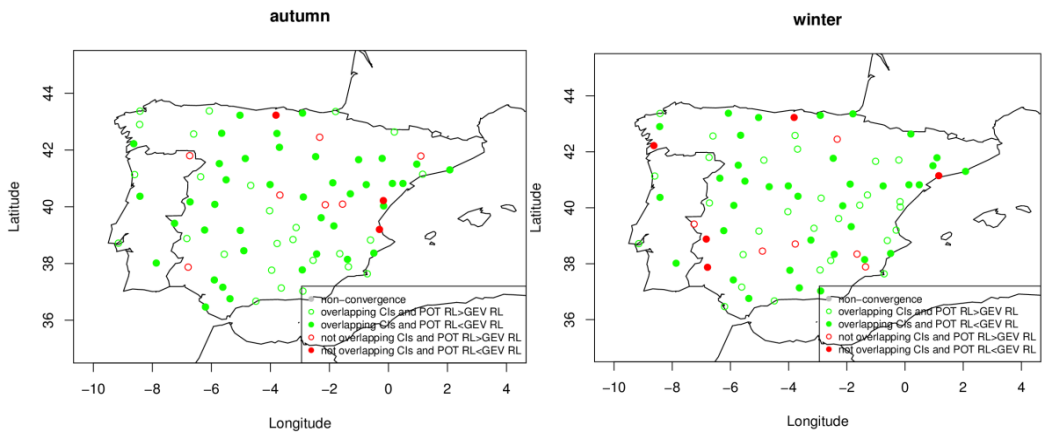

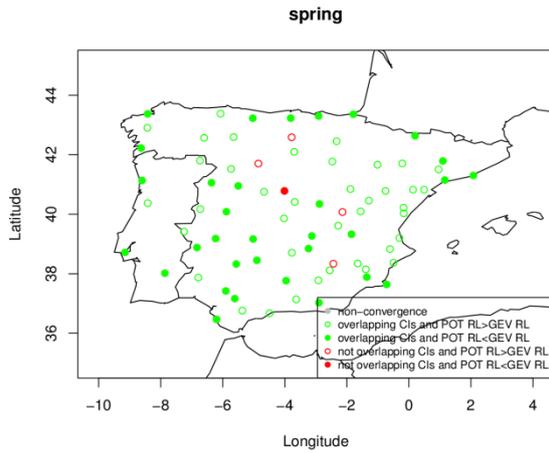

Figure 6. Spatial distribution of the 20-year RLs in 2020 obtained through the extrapolation of the location and scale parameters using GEV that lie or do not lie inside the CI of the 20-year RLs obtained through the extrapolation of the scale parameter using GPD.

Table 2. Degree of the parameters for GEV (location and scale) and POT (scale) for autumn. 0 means no trend and 1 linear trend.

| Code | 2 | 6 | 14 | 20 | 37 | 38 | 61 | 8 | 64 | 72 |
|---|---|---|---|---|---|---|---|---|---|---|
| Dmu-GEV | 1 | 0 | 0 | 0 | 0 | 0 | 0 | 0 | 0 | 1 |
| Dsig-GEV | 0 | 0 | 0 | 1 | 0 | 1 | 0 | 0 | 1 | 0 |
| Dsig-POT | 0 | 1 | 0 | 0 | 0 | 0 | 1 | 1 | 0 | 1 |
| | Red.open (not overlapping and POT RL>GEV RL) | | | | | | | Red.filled (not overlapping and POT RL<GEV RL) | | |

As a varying threshold for POT was considered, this is roughly equivalent to consider a linear trend in the location parameter ($\mu$) for GEV. Thus, those situations where $\mu$ shows linear trend and $\sigma$ for GEV and GPD shows no trend should be similar. But those gauges that disagree (like "agr"), may be due to the sampling because lower values are considered using BM approach. Figure 7 shows the behavior for both approaches for the gauge "agr" in autumn. The decrease in location is higher than the decrease in the threshold due to the lower maximum values at the last part of the time series. Even considering the maximum with BM approach (red circles) lower than the time-varying threshold, most of them are located at the end of the observed period and correspond to years showing none maximum value with POT technique. This leads to a lower value of 20-year RL in 2020 using GEV than using POT when extrapolating the parameters of the extreme value distribution.

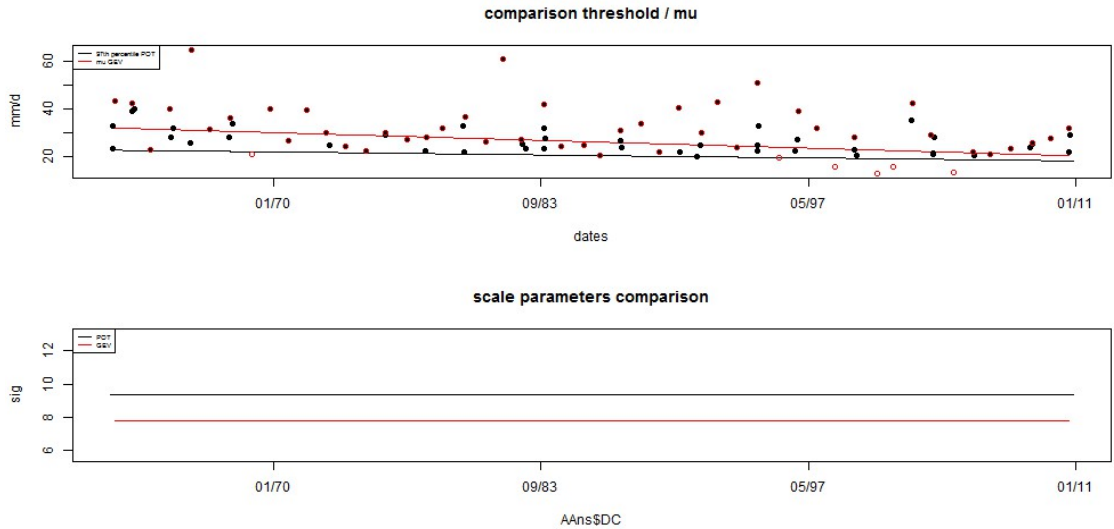

Figure 7. GEV-POT comparison for the gauge "agr" in autumn.

Table 3. Degree of the parameters for GEV (location and scale) and POT (scale) for winter. 0 means no trend and 1 linear trend.

| Code | cal | dto | eca | lib | log | val | alr | bad | reu | vig | vil |
|---|---|---|---|---|---|---|---|---|---|---|---|
| Dmu-GEV | 1 | 1 | 0 | 1 | 0 | 0 | 0 | 0 | 0 | 0 | 0 |
| Dsig-GEV | 0 | 0 | 0 | 0 | 0 | 0 | 1 | 0 | 0 | 0 | 0 |
| Dsig-POT | 0 | 0 | 0 | 1 | 0 | 1 | 0 | 0 | 1 | 1 | 0 |
| | Red.open (not overlapping and POT RL>GEV RL) | | | | | | Red.filled (not overlapping and POT RL<GEV RL) | | | | |

Figure 8 shows the comparison for both approaches using the gauge "dto" in winter. Again, the decrease in the location parameter is higher than the decrease in the threshold due to the lower maximum values mainly at the second half of the time series. This leads to a lower value of Z20 in 2020 using GEV than using POT. The same happens for the gauge "cal".

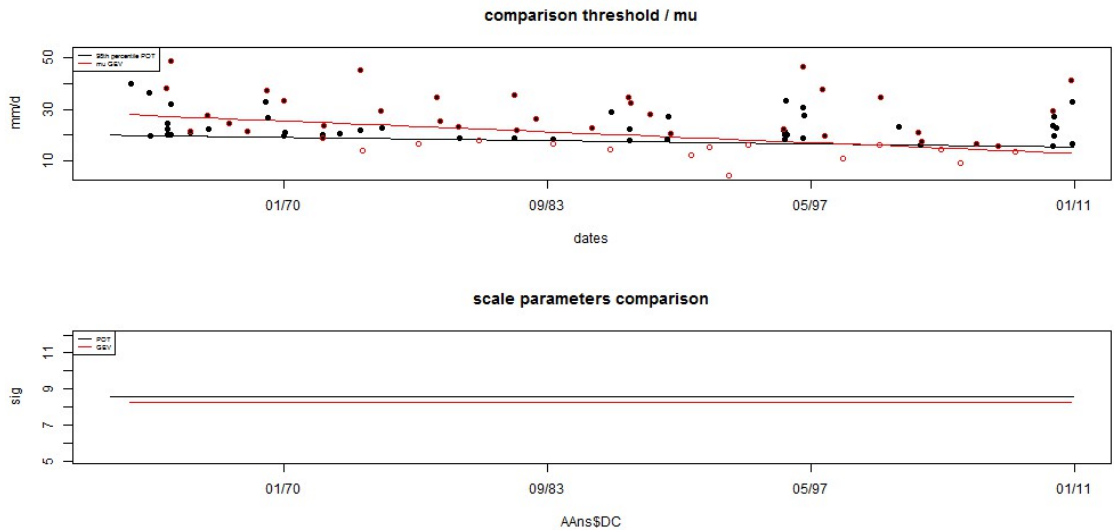

Figure 8. GEV-POT comparison for the gauge "dto" in winter.

Table 4. Degree of the parameters for GEV (location and scale) and POT (scale) for spring. 0 means no trend and 1 linear trend.

| Code | agu | cue | mon | vdl | nav |
|---|---|---|---|---|---|
| Dmu-GEV | 0 | 0 | 0 | 0 | 0 |
| Dsig-GEV | 1 | 1 | 0 | 0 | 0 |
| Dsig-POT | 0 | 0 | 1 | 1 | 1 |
|  | Red.open (not overlapping and POT RL>GEV RL) | | | Red.filled (not overlapping and POT RL<GEV RL) | |

4.3.2. Using the stationarity test.

The next step is to estimate the 20-year RLs in 2020 obtained by extrapolating the linear trends in the daily mean and standard deviation of the amount of rain for rainy days and the number of rainy days [12]. Previously, the hypothesis that the non-parametric temporal evolution is essentially linked to the evolutions of the mean and the variance, the methodological approach mentioned in [12], was used to test for the stationarity of the extremes of the standardized residuals of the time series. Table 5 shows the percentage of the gauges that verified this stationarity at a 90% confidence level totally. The stationarity test is rather well verified for both methodologies: for BM approach, in autumn and spring 87% of the gauges satisfied the test while winter shows a lower percentage of 78%; for POT approach, the percentage of gauges that verified the test is higher than for BM, being 95% for autumn and spring, and 92% for winter.

Table 5. Percentage of the stations that satisfy completely the stationarity of the extremes of the residuals for the three seasons considered using both methodologies.

|  | GEV | POT |
|---|---|---|
| AUTUMN | 87% | 95% |
| WINTER | 78% | 92% |
| SPRING | 87% | 95% |

Then, once tested that the stationarity test is rather well verified for both methodologies, the 20-year RLs in 2020 were estimated. Figure 9 shows the spatial distribution of the stations with the RL obtained with POT inside the CI of the RL obtained with GEV. The comparison produced similar results for both methodologies obtaining overlapping CIs for all the gauges in winter and all except for one in autumn and spring.

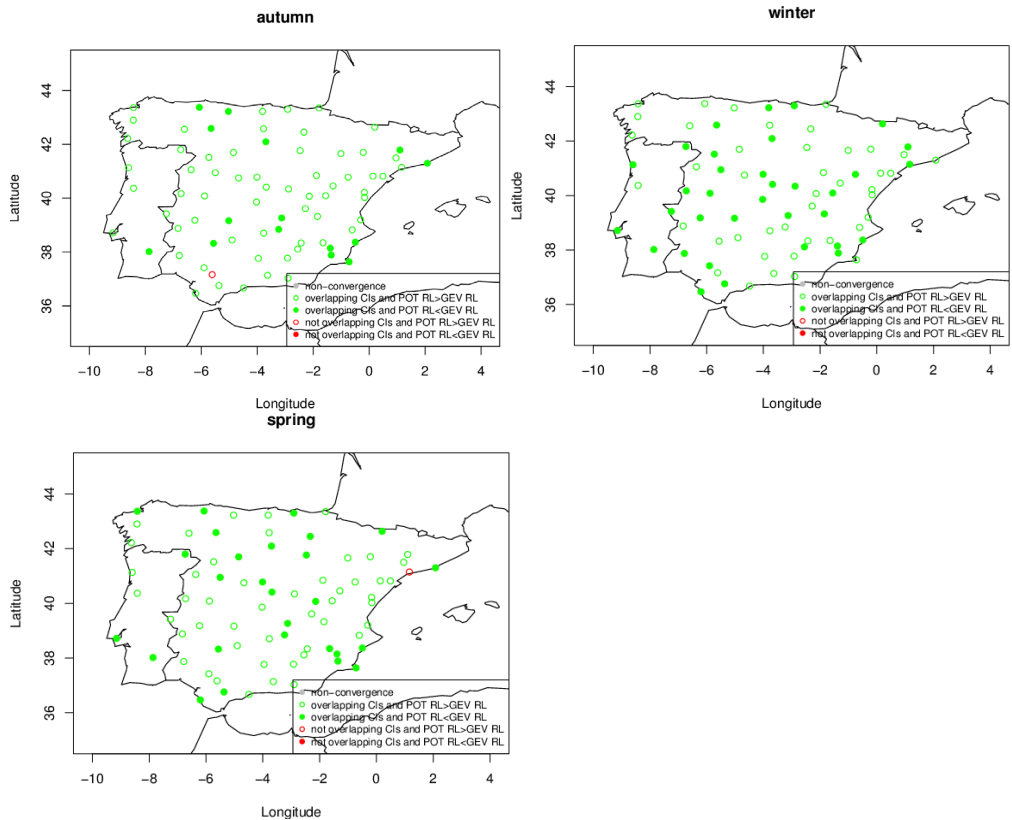

Figure 9. Spatial distribution of the 20-year RLs in 2020 obtained through the stationarity test for POT that lie or do not lie inside the CI of the 20-year RLs in 2020 obtained for BM.

It should be noted that now trends in the parameters are not extrapolated, but trends in mean and variance, which are the same for both approaches for mean but not for variances as explained below. The differences in Z20 results thus can come from the variance trends and/or from the selected high values of $Y(t)$ (annual maxima or excesses of the 95th percentile, which may lead to different extremes for $Y(t)$). Besides, we take into account the proportion of rainy days with POT, not with GEV, and then, the computation of the extremes of $X(t)$ from those of $Y(t)$ are different. Looking in greater depth at the exceptions, both gauges are analyzed

1) There is one gauge in autumn (red circle) "mfr" with Z20-POT outside the CI of Z20-BM:

| Code: mfr | Z20 |
|---|---|
| BM | 73.39 [63.96; 80.43] |
| POT | 81.04 [64.41; 86.03] |

This gauge shows negative (positive) trend for variance using BM (POT) leading to different results for Z20 when extrapolating these trends and also, there is positive trend in the number of rainy days. But the variance is different for both methods because for BM all days were used and not only rainy days that were used for POT approach. This is another drawback of the BM approach: mixing rainy and non-rainy days mixes 2 different processes and then, the variance mixes them too. Figure 10 shows the comparison for both approaches using the gauge "mfr" in autumn (two top panels) and the evolution of mean, standard deviation and number of rainy days in three bottom panels (seasonal values in black, linear trends in red). It is highlighted the importance of consider rainy days only, because a positive trend in variance of rain of rainy days and in the number of rainy days can result in negative trend in variance when all days are considered.

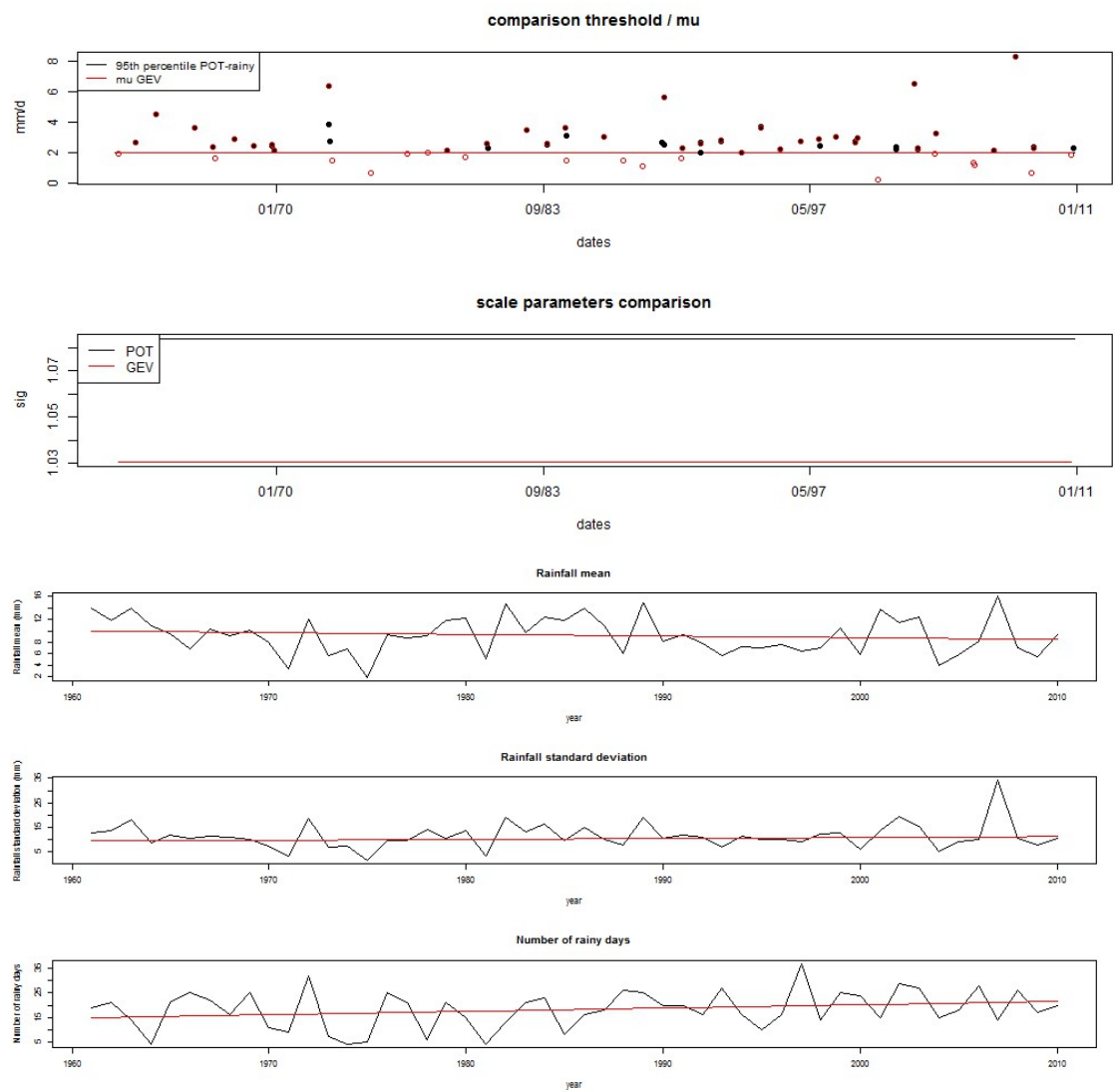

Figure 10. Comparison GEV-POT for gauge "mfr" in autumn but considering Y (two top panels) and evolution of mean, standard deviation and number of rainy days (seasonal values in black, linear trends in red) in the three bottom panels .

2) There is another gauge for spring (red circle) with Z20-POT outside the CI of Z20-GEV. It shows positive trend for variance for both methods but a positive trend in the number of rainy days producing a higher Z20 when using POT technique.

| Code: reu | Z20 |
|---|---|
| BM | 44.80 [33.64; 51.11] |
| POT | 51.58 [36.46; 59.10] |

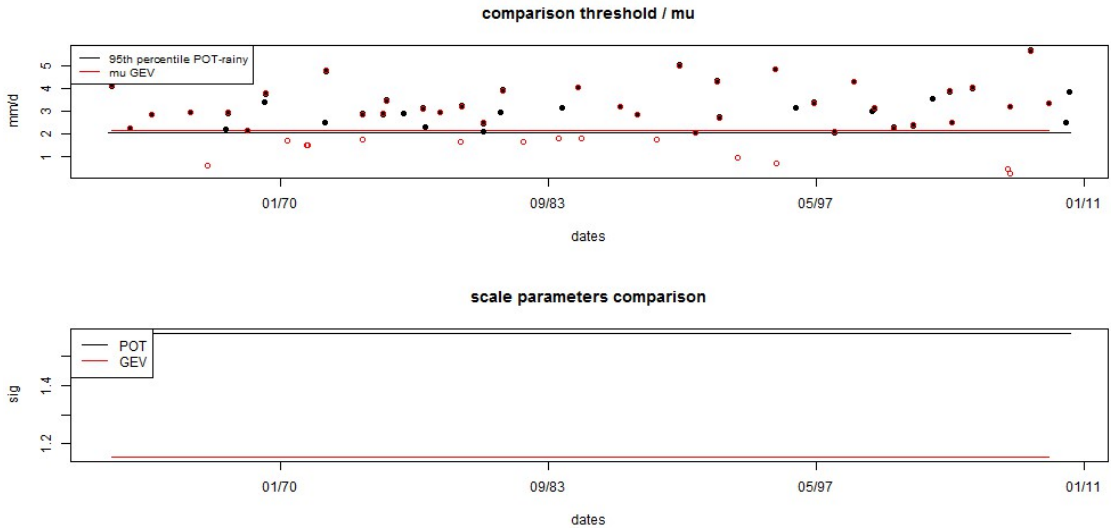

Figure 11. Comparison GEV-POT for gauge "reu" in spring but considering Y.

### 4.4. Changes in future 20-year RLs.

For both approaches, block maxima and POT, the spatial distribution of the differences between the 20-year RLs obtained in 2020 using the last approach and the corresponding in the present period are shown in Figure 12. The changes for block maxima are on the left, the POT ones on the right, with blue (red) meaning decreasing (increasing) values of the 20-year RLs in 2020. The scale used for both approaches is the same. Generally, for the three seasons considered both approaches show similar results although there are some remarks. Autumn shows an increase in western IP and a decrease in the east although the southeast shows an increase too. The increase in the northeast with BM (left) is higher than with POT. In winter, most of the IP show a decrease in the 20-year RLs in 2020 and only a small area in the southeast shows increasing RLs, being more noted with POT approach. However, most of these changes are not significant because the CIs overlap for most of the gauges. This is important because, the estimation of return levels is highly uncertain, especially in the climate change context.

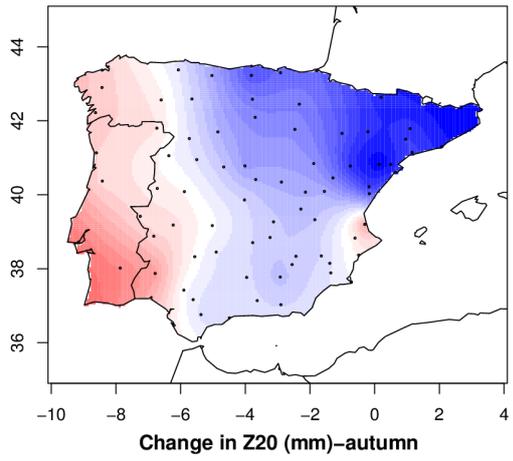
**Change in Z20 (mm)–autumn**

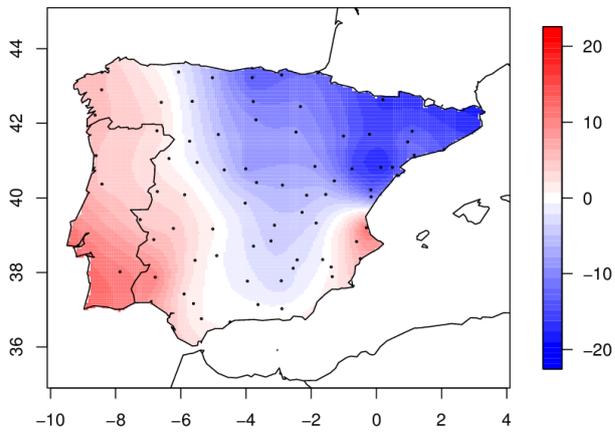
**Change in Z20 (mm)–autumn**

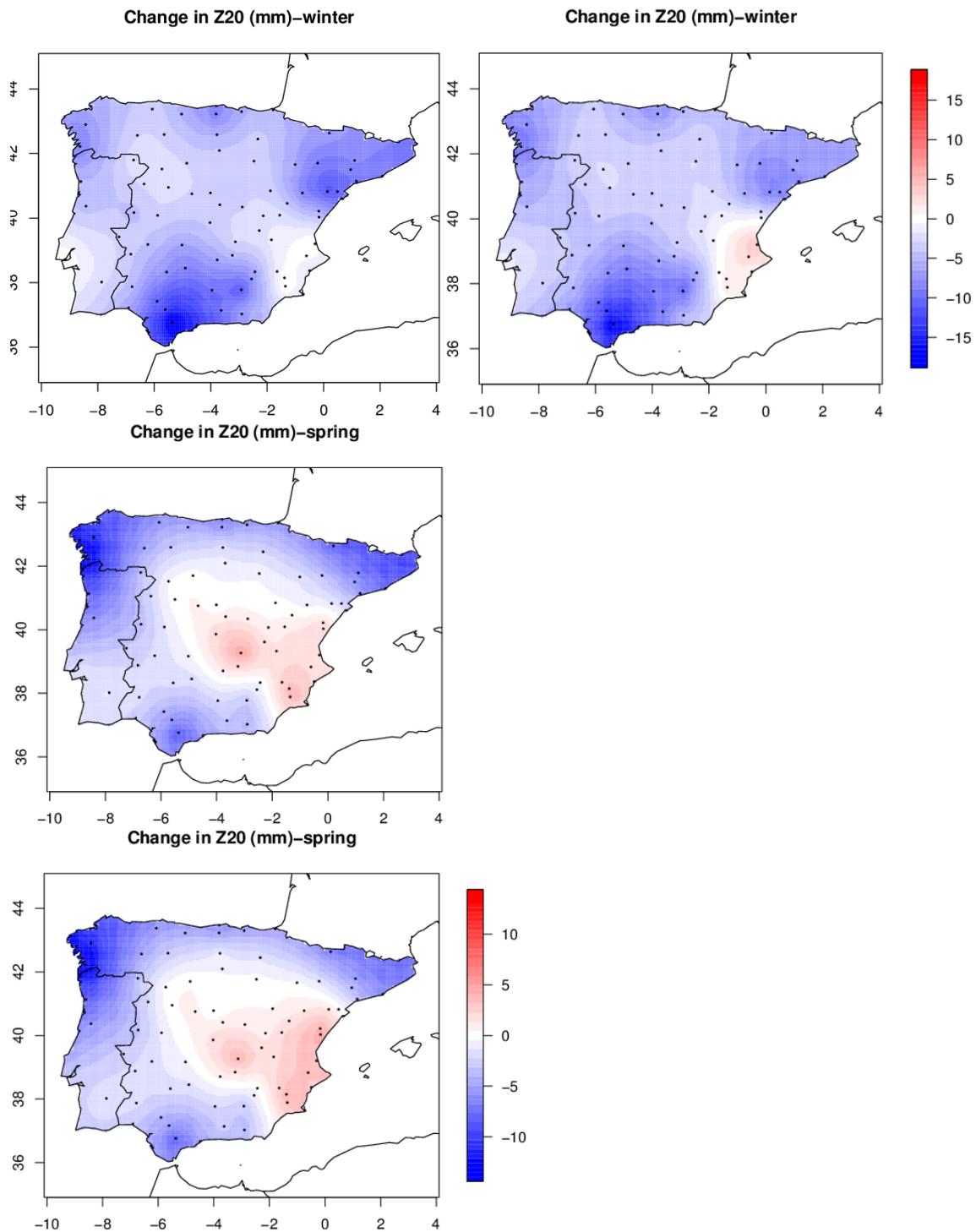

Figure 12. Spatial distribution of the differences between the present and the future for the 20-year RLs (mm) for each season considered for BM (left) and POT (right).

## 5. Conclusions

A study of the EVT to estimate non-stationary return levels of extreme rainfall for present and near future has been developed for the Iberian Peninsula. A complete dataset of 76 gauges for the period 1961-2010 was used. The two widely approaches in EVT, block maxima and peaks-over-threshold were used in order estimate the 20-year RLs in present time and the expected in 2020.

Two approaches to computing future rainfall RLs with POT were studied. In the first, trends in the extremes considering all the days were identified, taking into account a time-varying threshold based on a linear quantile regression and, when appropriate, a trend in the GPD scale parameter. Then, in the second, we calculated the RLs considering only the rainy days, examining the impact of evolutions of the mean and variance and of the number of rainy days. In this second case, we applied a novel adaptation of a stationarity test to rainfall that had been designed and used for temperature time series, finding that it was indeed satisfied for the majority of the gauges for all three seasons considered.

The estimation of the 20-year RLs for present time considering, all days (rainy and non-rainy) and rainy days only lead to similar results for the two approaches used. Only a few number of gauges show different RLs for BM and POT that have been highlighted in the previous section. As aforementioned, when using BM the asymptotic is weakened because in the rainfall time series in such a place of the Iberian Peninsula, there are lots of non-rainy days and the maximum of a block sometimes is not a real maximum.

Besides, the 20-year RLs expected in 2020 were estimated. The main objective was to compare the two methodologies of the EVT, block maxima and peaks-over-threshold. But there are some exceptions that confirm that peaks-over-threshold is a better methodology to estimate RLs because block maxima showed some gaps:

- Considering the decreasing trend in rainfall for different seasons [10,12], for several gauges, the BM approach considers maximum values at the second half of the study period that are not really a maximum. When comparing this selection with the corresponding to POT, as BM forces to take a maximum for each season, these low values lead to lower values of the RLs.
- Gauges showing a trend in the number of rainy days lead to different values in the RLs obtained with POT when using rainy days only than using all days with BM approach.

To summarize the comparison of both EVT methodologies, BM is less reliable than POT because fixed blocks lead to the selection of non extreme values. Besides, from the study of the changes in the near future RLs over the Iberian Peninsula, autumn becomes the season of heaviest rainfall, rather than winter in last recent decades, for some regions due to the increase in the RLs for the western IP.

Acknowledgments: Thanks are due to the Spanish Weather Agency (Agencia Estatal de Meteorología: www.aemet.es) for providing the daily rainfall time series used in this study. This work was partially supported by the Junta de Extremadura – FEDER Funds (IB16063) and Junta de Extremadura – FEDER Funds (GR15137).